\documentclass[a4paper,11pt]{article}
\pdfoutput=1
\usepackage{float}
\usepackage{graphicx}
\usepackage{dcolumn}
\usepackage{bm}
\usepackage{latexsym}
\usepackage{subfig}
\usepackage{amsmath}
\usepackage{jcappub} 

\usepackage[T1]{fontenc} 
\newcommand{\apar}{\alpha_\parallel}
\newcommand{\aperp}{\alpha_\perp}

\title{\boldmath Holographic Dark Energy with Cosmological Constant}

\author[a,b,c]{Yazhou Hu,}
\author[c,a,1]{Miao Li,\note{Corresponding author.}}
\author[a,b,c]{Nan Li,}
\author[a,b,c,1]{Zhenhui Zhang}

\affiliation[a]{State Key Laboratory of Theoretical Physics, Institute of Theoretical Physics, Chinese Academy of Sciences, Beijing, 100190}
\affiliation[b]{Kavli Institute for Theoretical Physics China, Chinese Academy of Sciences, Beijing, 100190}
\affiliation[c]{School of Astronomy and Space Science, Sun Yat-Sen University, Guangzhou 510275, People's Republic of China}

\emailAdd{asiahu@itp.ac.cn}
\emailAdd{mli@itp.ac.cn}
\emailAdd{linan@itp.ac.cn}
\emailAdd{zhangzhh@mail.ustc.edu.cn}

\abstract{Inspired by the multiverse scenario, we study a heterotic dark energy model in which
there are two parts, the first being the cosmological constant and the second being the holographic
dark energy, thus this model is named the $\Lambda$HDE model.
By studying the $\Lambda$HDE model theoretically, we find that the parameters $d$ and $\Omega_{hde}$
are divided into a few domains in which the fate of the universe is quite different.
We investigate dynamical behaviors of this model, and especially the future evolution of the universe.
We perform fitting analysis on the cosmological parameters in the $\Lambda$HDE model by using the recent observational data.
We find the model yields $\chi^2_{\rm min}=426.27$ when constrained by $\rm Planck+SNLS3+BAO+HST$,
comparable to the results of the HDE model (428.20) and the concordant $\Lambda$CDM model (431.35).
At 68.3\% CL, we obtain $-0.07<\Omega_{\Lambda0}<0.68$ and correspondingly  $0.04<\Omega_{hde0}<0.79$,
implying at present there is considerable degeneracy between the holographic dark energy and
cosmological constant components in the $\Lambda$HDE model.}

\begin{document}
\maketitle
\flushbottom

\section{Introduction}

The holographic dark energy model (HDE) \cite{Cohen,Li1} was motivated by the holographic principle \cite{Holography}, as one of promising models to solve the nature of dark energy \cite{Riess,DEReview}.
The basic idea behind the HDE is that our universe is a sense finite and can be described by a two dimensional spherical holographic screen, thus there must be finite size effects, and one of these effects is the contribution to the zero point energy, depending on the size of the screen.
Parametrically, this contribution assumes the form
\begin{equation}\label{eq:rhoHDE}
 \rho_{hde}=3d^2M^2_{pl}L^{-2},
\end{equation}
where $d$ is a dimensionless parameter to be determined by experiments, $M_{pl}$ is the reduced Planck mass, and $L$ denotes the size of holographic screen.
For convenience, we work in the natural units, where $\hbar=c=1$.
In \cite{Li1}, one of the present authors suggested to choose the future event horizon of the universe as the size of the holographic screen, given by
\begin{equation}\label{eq:Rh}
 R_h=a\int_t^{+\infty}\frac{dt}{a}.
\end{equation}
This choice not only gives a reasonable value for the dark energy density, but also leads to an accelerated expansion.

The holographic dark energy (HDE) model based on Eq.~(\ref{eq:rhoHDE}) and Eq.~(\ref{eq:Rh}) has proven to be a promising dark energy candidate.
In the original paper \cite{Li1}, Li showed that the HDE  can explain the coincidence problem.
In \cite{HDEstable}, it is proven that the model is perturbatively stable.
Other studies show that the model is in good agreement with the current cosmological observations \cite{HDEObserv}.
Thus, the HDE model becomes one of the most competitive and popular dark energy candidates, and attracts a lot of interests \cite{HDEworks}.

It remains quite a mystery that to date all the papers on the HDE assume that dark energy is dominated by the
HDE given by Eq.(\ref{eq:rhoHDE}).
In retrospect, this can be explained only by the reason that all the
authors believed that the universe dominated by the HDE is the unique universe thus the form of dark energy is also unique.
In the multiverse scenario, however, our observable universe is only one out of numerous universes, and the cosmological constant is one of the physics constants varying from one universe to another. Thus, it is not
reasonable to simply assume it be vanishing.

Of cause, the modern multivese scenario was motivated by the problem of dark energy. One of the present authors (ML) has
recently been converted into a believer of the multiverse scenario by a quite different problem, namely the Fermi
paradox. He now believes that the correct answer to the Fermi paradox is that our human being is the only
intelligent being in our galaxy, possibly the only intelligent being in our universe, since it is really
very difficult for an intelligent being to appear, the only reason for us to appear is that our universe
is one out of numerous universe and it just happens that our universe is lucky enough, this is an anthropic answer
to the Fermi paradox.

If for whatever reason that the multiverse scenario is true, then it is natural that the cosmological constant indeed is a constant to be determined by observations.
On the other hand, the HDE on a general ground must be present too, according to the holographic principle.
Thus, dark energy must consist of two parts, the first is a constant, the second is of the finite size effect.
Indeed, in a calculation of the photon contribution to the zero point energy \cite{Casimir}, it is found that in addition to the usual UV divergent part, there is a second divergent part proportional to $L^{-2}$ where $L$ is the radius of the de Sitter space.
The usual quartically divergent part can be absorbed into the cosmological constant, the second part is the same form of the HDE.
Thus, in general
\begin{equation}\label{eq:Rh2}
 \rho_{IR} = \Lambda + bM_{pl}^2 L^{-2}  + b_1L^{-4}+\dots ~.
\end{equation}
It just happens that in our universe, the first term and the second term are comparable, and the third term is much smaller and can be neglected completely.

In this paper, we shall study this heterotic model of dark energy, in which there are the cosmological constant and the HDE.
Thus we have
\begin{equation}\label{eq:lhde}
 \rho_{de} = \rho_{\Lambda} + \rho_{hde}.
\end{equation}
We shall name this model $\Lambda$HDE model.
This model raises interesting theoretically questions: Is the CC positive or negative?
If it is positive, whether it is greater or smaller than the HDE?
Is the future of our universe dominated by the CC or by HDE?
Will the big rip happen or not?
In the rest of this paper, we shall try to answer these questions.

This paper is organized as follows.
We study the $\Lambda$HDE model theoretically in Sect.II, and find that the parameters $d$ and $\Omega_{hde}$
are divided into a few domains in which the fate of the universe is quite different.
In Sect.III, we fit the model to the combined $\rm Planck+SNLS3+BAO+HST$ and $\rm Planck+SNLS3+BAO+HST+SDSS-Ly\alpha$ datasets, and present the fittings results in Sec. IV.
Many interesting issues, including the ratio of HDE and the equation of state (EoS) of the $\Lambda$HDE are discussed.
Some concluding remarks are given in Sec. V.
In this work, we assume today's scale factor $a_0 = 1$, so the redshift $z$ satisfies $z = 1/a - 1$.
The subscript ``0'' indicates the present value of the corresponding quantity unless otherwise specified.

\section{$\Lambda$HDE Model: Theoretical  Analysis }

In this section, we will write down the basic equations for the $\Lambda$HDE model in a non-flat universe and study the fate of the universe in this model.

\subsection{HDE with Cosmological Constant}

In a spatially non-flat Friedmann-Robertson-Walker universe, the Friedmann equation can be written as
\begin{equation}
\label{eq:FE1} 3M_{pl}^2 H^2=\rho_{dm}+\rho_b+\rho_r+\rho_k+\rho_{de},
\end{equation}
where $\rho_k=-3M_{pl}^2\frac{k}{a^2}$ is the effective energy density of
the curvature component.
In the $\Lambda$HDE model, the dark energy density is
\begin{equation}
\label{eq:rholhde}
\rho_{de} =\rho_{\Lambda}+\rho_{hde}=M_{pl}^2\Lambda +3M_{pl}^2d^2R_h^{-2}.
\end{equation}
For convenience, we define the fractional energy densities of the various components, i.e.,
\begin{equation}
\label{eq:DefOmega_k} \Omega_k={-k \over
H^2a^2},\ \ \ \ \
\Omega_{de}={\rho_{de} \over \rho_c},\ \ \ \ \
\Omega_{hde}={\rho_{hde} \over \rho_c},\ \ \ \ \
\Omega_{\Lambda}={\rho_{\Lambda} \over \rho_c},\ \ \ \ \
\Omega_{dm}={\rho_{dm} \over \rho_c},\ \ \ \
\Omega_{b}={\rho_{b} \over \rho_c},\ \ \ \
\Omega_r ={\rho_r \over  \rho_c},
\end{equation}
where $\rho_c=3M_{pl}^2H^2$ is the critical density of the universe.
The subscripts, ``$k$'', ``$de$'',  ``$hde$'', ``$\Lambda$'', ``$dm$'', ``$b$'' and ``$r$'' represent curvature,  total dark energy,
holographic dark energy, cosmological constant, dark matter, baryon and radiation, respectively. By definition, we have
\begin{equation}
\label{eq:AllOmega} \Omega_{hde}+\Omega_{\Lambda}+\Omega_{dm}+\Omega_b+\Omega_r+\Omega_k=1.
\end{equation}

The energy conservation equations for the components in the universe take the forms
\begin{equation}
\label{eq:IHDEeq1}\dot\rho_{hde}+3H(\rho_{hde}+p_{hde})=0,
\end{equation}
\begin{equation}
\label{eq:lambaeq}\dot\rho_{\Lambda}=0,\ \ \ \ \
\dot\rho_{dm}+3H\rho_{dm}=0,\ \ \ \ \
\dot\rho_b+3H\rho_b=0,\ \ \ \ \
\dot\rho_r+4H\rho_r=0,\ \ \ \ \
\dot\rho_k+2H\rho_k=0.
\end{equation}
Combining Eq.~(\ref{eq:IHDEeq1}) and Eq.~(\ref{eq:lambaeq})
together, we can obtain the form of $p_{hde}$,
\begin{equation}\label{eq:pde}
p_{hde}=-\frac{2}{3}\frac{\dot H}{H^2}\rho_c-\rho_c-{1\over3}\rho_r+{1\over3}\rho_k+\rho_{\Lambda}.
\end{equation}
Substituting $p_{hde}$ into Eq. (\ref{eq:IHDEeq1}), follow the similar procedure in Ref.~\cite{zhang2012},
we get a differential equation of $\dot H$ and $\dot \Omega_{hde}$:
\begin{equation}
\label{eq:OH2} 2(\Omega_{hde}-1){\dot H\over
H}+\dot\Omega_{hde}+H(3\Omega_{hde}-3+3\Omega_{\Lambda}+\Omega_k-\Omega_r)=0.
\end{equation}
From the energy density of the HDE in Eq.~(\ref{eq:rhoHDE}), we have
\begin{equation}
\label{ea:L0} L={d \over H\sqrt{\Omega_{hde}}}.
\end{equation}
Following Ref.~\cite{holode1}, in a non-flat universe, the IR cut-off length scale $L$ takes the form
\begin{equation}
\label{eq:L1} L=ar(t),
\end{equation}
and $r(t)$ satisfies
\begin{equation}
\label{eq:r(t)1}  \int_0^{r(t)} {dr \over
\sqrt{1-kr^2}}=\int_t^{+\infty}{dt\over a(t)}.
\end{equation}
By carrying out the integration, we find
\begin{equation}
\label{eq:r(t)2} r(t)={1\over\sqrt{\left|k\right|}}{\rm sinn}
\Big(\sqrt{\left|k\right|}\int_t^{+\infty} {dt \over a}\Big)={1\over\sqrt{\left|k\right|}}{\rm sinn}
\Big(\sqrt{\left|k\right|}\int_{a(t)}^{+\infty} {da \over {Ha^2}}\Big),
\end{equation}
where the function sinn(x) is defined as
\begin{displaymath}
  {\rm sinn}(x)=\left\{ \begin{array}{ll}
      {\rm sin}(x)  & \textrm{$k>0$}; \\
      x             & \textrm{$k=0$}; \\
      {\rm sinh}(x) & \textrm{$k<0$}. \\
    \end{array} \right.
\end{displaymath}
Equation (\ref{eq:L1}) leads to another equation about $r(t)$, namely,
\begin{equation}
\label{eq:r(t)3.0} r(t)={L\over
a}={d\over\sqrt{\Omega_{hde}}Ha}.
\end{equation}
Combining Eqs. (\ref{eq:r(t)2}) and (\ref{eq:r(t)3.0}) yields
\begin{equation}
\label{eq:OL1.1}\sqrt{\left|k\right|}\int_t^{+\infty}{dt\over
a}={\rm arcsinn}{\sqrt{\left|k\right|}d\over \sqrt{\Omega_{hde}}aH}.
\end{equation}
Taking derivative of Eq.~(\ref{eq:OL1.1}) with respect to $t$, one gets
\begin{equation}
\label{eq:OL1.3}
{\dot\Omega_{hde}\over2\Omega_{hde}}+H+{\dot H\over H}=\sqrt{{\Omega_{hde}H^2\over d^2}-{k\over a^2}}.
\end{equation}

\subsection{Evolution Equations of $E(z)$ and $\Omega_{hde}(z)$}

Combining Eq.~(\ref{eq:OH2}) with Eq.~(\ref{eq:OL1.3}), we eventually obtain the following two equations governing the dynamical evolution of the $\Lambda$HDE model in a non-flat universe,
\begin{equation}
\label{eq:OH3}{1\over E(z)}{dE(z) \over dz}
=-{\Omega_{hde}\over
1+z}\left({3\Omega_{\Lambda}+\Omega_k-\Omega_r-3\over2\Omega_{hde}}+{1\over2}+\sqrt{{\Omega_{hde}\over
d^2}+\Omega_k} \right),
\end{equation}
\begin{equation}
\label{eq:OH4}
{d\Omega_{hde}\over dz}=
-{2\Omega_{hde}(1-\Omega_{hde})\over 1+z}\left(\sqrt{{\Omega_{hde}\over
d^2}+\Omega_k}+{1\over2}-{3\Omega_{\Lambda}+\Omega_k-\Omega_r\over 2(1-\Omega_{hde})}\right),
\end{equation}
where $E(z)\equiv H(z)/H_0$ is the dimensionless Hubble expansion rate.
Notice that we have
\begin{equation}
\Omega_k(z)=\frac{\Omega_{k0}(1+z)^2}{E(z)^2},\ \  \Omega_r(z)=\frac{\Omega_{r0}(1+z)^4}{E(z)^2},\ \
\Omega_{\Lambda}(z)=\frac{\Omega_{\Lambda0}}{E(z)^2},\ \
\Omega_b(z)=\frac{\Omega_{b0}(1+z)^3}{E(z)^2},
\end{equation}
and the fractional density of dark matter is given by $\Omega_{dm}(z)=1-\Omega_k(z)-\Omega_{hde}(z)-\Omega_{\Lambda}(z)-\Omega_r(z)-\Omega_b(z)$.
Equations (\ref{eq:OH3}) and (\ref{eq:OH4}) can be solved numerically and will be used in the data analysis procedure.

\subsection{Dark Energy Equation of State}\label{subsec:DEEoS}

The EoS of the HDE takes the form \cite{NewHDE}
\begin{equation}
w_{\rm hde}\equiv\frac{p_{\rm hde}}{\rho_{\rm hde}}=-{1\over3}-{2\over3}\sqrt{{\Omega_{hde}\over d^2}+\Omega_k}.
\end{equation}
So according to the partial pressure law, the  EoS of the total dark energy is
\begin{equation}\label{eq:eosde}
w_{\rm de}\equiv\frac{p_{\rm de}}{\rho_{\rm de}}=\frac{p_{\rm hde}+p_{\rm \Lambda}}{\rho_{\rm hde}+\rho_{\rm \Lambda}}=\frac{w_{hde}\Omega_{hde}-\Omega_{\Lambda}}{\Omega_{hde}+\Omega_{\Lambda}}.
\end{equation}
Obviously the property of $w_{\rm de}$ is closely related to values of $d$ and $\Omega_{\Lambda}$.

\subsection{The Fate of the Universe in $\Lambda$HDE Model}

For convenience, we transform Eq.~(\ref{eq:FE1}) into the following form
\begin{equation}
(1-\Omega_{hde})H^{2} =\Omega_{m0}H^{2}_{0}a^{-3}+\Omega_{r0}H^{2}_{0}a^{-4}+\Omega_{k0}H^{2}_{0}a^{-2}+\Omega_{\Lambda0}H^{2}_{0},
\end{equation}
and define
\begin{equation}
\label{eq:fa} f(a)\equiv\Omega_{m0}H^{2}_{0}a^{-1}+\Omega_{r0}H^{2}_{0}a^{-2}+\Omega_{k0}H^{2}_{0}+\Omega_{\Lambda0}H^{2}_{0}a^{2}.
\end{equation}
Let $x\equiv\log a$, then we obtain
\begin{eqnarray}
  H &=& \sqrt{\frac{f(a)}{a^2(1-\Omega_{hde})}} \label{H},
  \label{main}
\end{eqnarray}

\begin{equation}
  \frac{d}{dx}\ln{\left|\frac{\Omega_{hde}}{1-\Omega_{hde}}\right|} + \frac{d}{dx}\ln{\left|f(a)\right|}
  =\frac{2}{d}\sqrt{\Omega_{hde}-(1-\Omega_{hde})\frac{kd^2}{f(a)}} \label{main0}.
\end{equation}

Eq.~(\ref{main0}) tells us how the HDE evolves with $a$. For simplicity, in this section we only study
the $k=0$ case.
Then Eq.~(\ref{main0}) becomes
\begin{equation}
  \frac{d}{dx}\ln{\left|\frac{\Omega_{hde}}{1-\Omega_{hde}}\right|} + \frac{d}{dx}\ln{\left|f(a)\right|}=\frac{2}{d}\sqrt{\Omega_{hde}}. \label{main}
\end{equation}
This equation can not be solved exactly. But for the purpose of studying the fate of the universe, we can introduce a good approximation.

During a period in which $f(a)$ is dominated by a single term on the right hand side of Eq.~(\ref{eq:fa}), we can use a constant $k_r$ to approximate $\frac{d}{dx}\ln{\left|f(a)\right|}$.
Then the Eq.~(\ref{main}) becomes
\begin{equation}
\frac{d}{dx}\ln{\left|\frac{\Omega_{hde}}{1-\Omega_{hde}}\right|} = \frac{2}{d}\sqrt{\Omega_{hde}} - k_r \label{maink}.
\end{equation}

It is easy to see that the cosmological constant term on the right hand side of Eq.~(\ref{eq:fa}) will dominate when scale factor $a$ evolves to infinity. Thus we get $k_r=2$ in this limit.
Fortunately, it can be proved that $a$ will always evolve to infinity.
The proof is given in the appendix.

With $k_r=2$, we can rewrite the Eq.~(\ref{maink}) and get its solution as follows:
\begin{eqnarray}
  \frac{d\Omega_{hde}}{dx} &=& \frac{2}{d}(\sqrt{\Omega_{hde}} - d)(1-\Omega_{hde})\Omega_{hde}, \label{main2}
  \\
  x+x_{1} &=& \frac{d}{2d-2} \ln{\left| 1-\sqrt{\Omega_{hde}} \right|} -\frac{1}{(d^2-1)}
  \ln{\left|\sqrt{\Omega_{hde}}-d \right|}-\frac{1}{2}\ln{\Omega_{hde}}+{} \nonumber \\
  & & {} +\frac{d}{2d+2}\ln{(1+\sqrt{\Omega_{hde}})}, \label{mainsol2}
\end{eqnarray}
where $x_1$ is the constant of integration.
\begin{figure}[htbp]
  \centering
  \subfloat[$0<\sqrt{\Omega_{hde0}}<1$]{
\label{figA}
  \begin{minipage}{90pt}
\setlength{\unitlength}{0.75cm}
\footnotesize
\begin{picture}(6,3.5)
  \put(0,2){\line(1,0){6}}
  \put(1,2){\line(0,1){1.4}}
  \put(2.3,2.7){\vector(-1,0){1.3}}
  \put(1,2){\circle*{0.2}}
  \put(3,2){\circle*{0.2}}
  \put(5,2){\circle*{0.2}}
  \put(4.9,1.5){$d$}
  \put(2.9,1.5){$1$}
  \put(0.9,1.5){$0$}
  \put(1.4,2.9){$\sqrt{\Omega_{hde}}$}
\end{picture}
\end{minipage}
}
\hspace{60pt}
\subfloat[$0<\sqrt{\Omega_{hde0}}<d$]{
\label{figB}
\begin{minipage}{90pt}
\setlength{\unitlength}{0.75cm}
\footnotesize
\begin{picture}(6,3.5)
  \put(0,2){\line(1,0){6}}
  \put(1,2){\line(0,1){1.4}}
  \put(2.3,2.7){\vector(-1,0){1.3}}
  \put(1,2){\circle*{0.2}}
  \put(3,2){\circle*{0.2}}
  \put(5,2){\circle*{0.2}}
  \put(4.9,1.5){$1$}
  \put(2.9,1.5){$d$}
  \put(0.9,1.5){$0$}
  \put(1.4,2.9){$\sqrt{\Omega_{hde}}$}
\end{picture}
\end{minipage}
}
\hspace{60pt}
\subfloat[{$d<\sqrt{\Omega_{hde0}}<1$}]{
\label{figC}
\begin{minipage}{90pt}
  \centering
\setlength{\unitlength}{0.75cm}
\footnotesize
\begin{picture}(6,3.5)
  \put(0,2){\line(1,0){6}}
  \put(5,2){\line(0,1){1.4}}
  \put(3.7,2.7){\vector(1,0){1.3}}
  \put(1,2){\circle*{0.2}}
  \put(3,2){\circle*{0.2}}
  \put(5,2){\circle*{0.2}}
  \put(4.9,1.5){$1$}
  \put(2.9,1.5){$d$}
  \put(0.9,1.5){$0$}
  \put(3.6,2.9){$\sqrt{\Omega_{hde}}$}
\end{picture}
\end{minipage}
}
\hspace{60pt}
\subfloat[$1<\sqrt{\Omega_{hde0}}$]{
\label{figD}
\begin{minipage}{90pt}
  \centering
\setlength{\unitlength}{0.75cm}
\footnotesize
\begin{picture}(6,3.5)
  \put(0,2){\line(1,0){6}}
  \put(3,2){\line(0,1){1.4}}
  \put(4.3,2.7){\vector(-1,0){1.3}}
  \put(1,2){\circle*{0.2}}
  \put(3,2){\circle*{0.2}}
  \put(2.9,1.5){$1$}
  \put(0.9,1.5){$d$}
  \put(3.6,2.9){$\sqrt{\Omega_{hde}}$}
\end{picture}
\end{minipage}
}
\hspace{60pt}
\subfloat[$d<\sqrt{\Omega_{hde0}}$]{
\label{figE}
\begin{minipage}{90pt}
  \centering
\setlength{\unitlength}{0.75cm}
\footnotesize
\begin{picture}(6,3.5)
  \put(0,2){\line(1,0){6}}
  \put(3,2){\line(0,1){1.4}}
  \put(4.3,2.7){\vector(-1,0){1.3}}
  \put(1,2){\circle*{0.2}}
  \put(3,2){\circle*{0.2}}
  \put(0.9,1.5){$1$}
  \put(2.9,1.5){$d$}
  \put(3.6,2.9){$\sqrt{\Omega_{hde}}$}
\end{picture}
\end{minipage}
}
\hspace{60pt}
\subfloat[$1<\sqrt{\Omega_{hde0}}<d$]{
\label{figF}
\begin{minipage}{90pt}
  \centering
\setlength{\unitlength}{0.75cm}
\footnotesize
\begin{picture}(6,3.5)
  \put(0,2){\line(1,0){6}}
  \put(5,2){\line(0,1){1.4}}
  \put(3.7,2.7){\vector(1,0){1.3}}
  \put(1,2){\circle*{0.2}}
  \put(3,2){\circle*{0.2}}
  \put(5,2){\circle*{0.2}}
  \put(4.9,1.5){$d$}
  \put(2.9,1.5){$1$}
  \put(0.9,1.5){$0$}
  \put(3.6,2.9){$\sqrt{\Omega_{hde}}$}
\end{picture}
\end{minipage}
}
\caption{\label{theory} The evolution of $\sqrt{\Omega_{hde}}$ with different initial conditions.}
\end{figure}
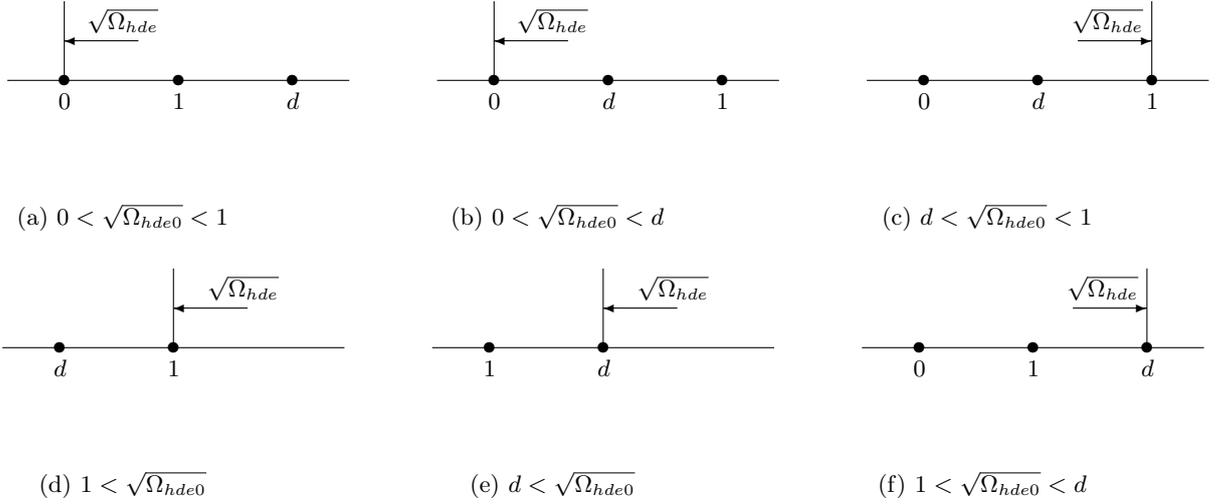
Then let us exhibit how the HDE evolves:

The first three pictures in Fig.\ref{theory} show how $\sqrt{\Omega_{hde}}$ evolves with the
initial condition $\sqrt{\Omega_{hde0}}<1$. In both picture(\ref{figA}) and picture(\ref{figB}),
$\sqrt{\Omega_{hde0}}<d$, the right hand of Eq.(\ref{main2}) is negative,
so $\sqrt{\Omega_{hde0}}$ will decrease with x.
Thus the right hand of Eq.(\ref{main2}) will always be negative.
So $\sqrt{\Omega_{hde0}}$ will keep decreasing to 0.
According to Eq.(\ref{mainsol2}) when $\sqrt{\Omega_{hde0}} \to 0^+, x \to +\infty$. Thus once
$\sqrt{\Omega_{hde0}}<1$, either $d<1$ or $d>1$, according to  Eq.(\ref{H}),
 $H$ will always tend to a constant,
which means space-time will be de Sitter in  the future. In picture (\ref{figC}),
$d<\sqrt{\Omega_{hde0}}<1$, the right hand of Eq.(\ref{main2}) is positive,
so $\sqrt{\Omega_{hde0}}$ will increase with x.
Thus the right hand of Eq.(\ref{main2}) will always be positive.
$\sqrt{\Omega_{hde0}}$ will keep increasing to 1,
according to Eq.(\ref{mainsol2}) when$ \sqrt{\Omega_{hde0}} \to 1^-, x \to +\infty$.
So this solution describes a HDE dominated universe with big rip.

The last three pictures in Fig.~\ref{theory} show how $\sqrt{\Omega_{hde}}$ evolves with the
initial condition $\sqrt{\Omega_{hde0}}>1$. In picture (\ref{figD}),
$\sqrt{\Omega_{hde0}}>d$, the right hand of  Eq.(\ref{main2}) is negative,
so when $d<1$, the $\sqrt{\Omega_{hde0}}$ will keep decreasing with x all the way up to 1.
According to  Eq.(\ref{mainsol2}) when$ \sqrt{\Omega_{hde0}} \to 1^+, x \to +\infty$.
So picture(\ref{figD}) describes a HDE dominated universe with big rip.
In picture(\ref{figE}) (or in picture(\ref{figF})), $d<\sqrt{\Omega_{hde0}}$ (or $d>\sqrt{\Omega_{hde0}}$)
the right hand of  Eq.(\ref{main2}) will be negative (positive),
so the $\sqrt{\Omega_{hde0}}$ will keep decreasing (or increasing) to d.
According to  Eq.(\ref{mainsol2}), when$ \sqrt{\Omega_{hde0}} \to d, x \to +\infty$.
So they both describe an universe whose space-time will be de Sitter in the future.

\section{The Observational Data and Methodology}

We explore cosmological constraints on the $\Lambda$HDE model with the most recent observational data.
For comparison, we will also present the fitting results of the original HDE and $\Lambda$CDM models.
Data used in our analysis include:
\begin{itemize}
\item
The $\rm SNLS3$ combined sample~\citep{SNLS3A,SNLS3B}, consisting of 472 SNIa,
combining the results of two light-curve fitting codes SiFTO \citep{SiFTO} and SALT2 \citep{SALT2}.
\footnote{It should be mentioned that, previous studies on the SNLS3 data sets \cite{WangWang13} found strong evidence for the redshift-dependence of color-luminosity parameter $\beta$, and
this conclusion has significant effects on parameter estimation of various cosmological models \cite{WLZ14,WWGZ14,WWZ14,Wang15}. In addition, different light-curve fitters of SNIa can also affect the results of cosmology-fits \cite{Bengochea11,Bengochea14,HLLW15}. But in this work, for simplicity, we just adopt the most mainstream recipe of processing SNLS3 data and do not consider the factors of time-varying $\beta$ and different light-curve fitters.}
We follow the procedure of \cite{Zhenhui Zhang} and perform a complete analysis of the systematic errors.
The $\chi^2$ function takes the form
\begin{equation}
\chi^2_{SNLS3}=\Delta \overrightarrow{\bf m}^T \cdot {\bf C}^{-1} \cdot \Delta \overrightarrow{\bf m},
\end{equation}
where {\bf C} is a $472 \times 472$ covariance matrix capturing the statistic and systematic uncertainties,
and $\Delta {\overrightarrow {\bf m}} = {\overrightarrow {\bf m}}_B - {\overrightarrow {\bf m}}_{\rm mod}$ is a vector of
model residuals of the SNIa sample,
with $m_B$ the rest-frame peak $B$ band magnitude of the SNIa and
$m_{\rm mod}$ the predicted magnitude of the SNIa, given by
\begin{equation}\label{SNchisq}
m_{\rm mod} = 5\log_{10} \mathcal{D}_L-\alpha(s-1)+\beta \mathcal{C} + \mathcal{M},
\end{equation}
where $\mathcal{D}_L$ is the Hubble-constant free luminosity distance,
the stretch $s$ is a measure of the shape of SN light-curve, $\mathcal{C}$ is the color measure for the SN,
and $\alpha$, $\beta$ are two nuisance parameters characterizing the stretch-luminosity and color-luminosity relationships, respectively.
Following \cite{SNLS3A}, we treat $\alpha$ and $\beta$ as free parameters of $\chi^2$ function.

\item
The Planck ``distance priors''provided in ~\cite{Yun Wang}, which are extracted from Planck first year \citep{PlanckA,PlanckB} observations.  The data include the baryon component $\omega_b\equiv \Omega_b h^2$, the ``acoustic scale''
$l_a\equiv{\pi r(z_*)/ r_{\rm s}(z_*)}$, and the ``shift parameter'' $R\equiv{\sqrt{\Omega_m H_0^2} \,r(z_*)}$,
where $z_*$ is the redshift to the photon-decoupling surface \citep{Hu:1995en},
$r(z_*)$ is our comoving distance to $z_*$,
and $r_{\rm s}(z_*)$ is the comoving sound horizon at $z_*$.
The distance priors provide an efficient summary of the CMB data in regards to dark energy constraints \citep{Li2011}.

\item
The BAO data including the measurement of $r_{\rm s}/D_{\rm V}$ at $z=0.106$ from 6dFGS ~\citep{Beutler},
the isotropic measurement of $D_{\rm V}/r_{\rm d}$ at $z=0.32$ from the BOSS DR11 LOWZ sample \citep{dr11},
the anisotropic measurement of $D_{\rm A}/r_{\rm d}$ and $Hr_{\rm d}$ at $z=0.57$ from the BOSS DR11 CMASS sample \citep{dr11},
and the improved measurements of $D_{\rm V}/{r_{\rm s}}$
at $z=0.44$, 0.60, 0.73 from the WiggleZ Dark Energy Survey \citep{impwigz}.
Here $r_{\rm d}$ is the comoving sound horizon at the ``drag'' epoch when the baryons are ``released'' from the drag of the photons \citep{Eisenstein:1997ik},
and $D_{\rm V}$ is a volume averaged distance indicator similar to the angular diameter distance $D_{\rm A}$ ~\citep{eisenstein.etc}.

\item
The Hubble constant measurement $H_0=73.8\pm 2.4 {\rm km/s/Mpc}$ from the WFC3 on the HST (Hubble Space Telescope) ~\citep{HSTWFC3}.

\item
The high-redshift BAO measurement from the Quasar-Ly$\alpha$-forest cross-correlation of the BOSS DR11 of SDSS-III~\citep{bao2014}, namely
$\apar^{0.7}\aperp^{0.3} = 1.025 \pm 0.021$;  $\apar$ and $\aperp$ are defined as
$\apar = \frac { \left[D_H(\bar z)/r_d\right] }{\left[D_H(\bar z)/r_d\right]_{\rm fid}}$,
$\aperp = \frac { \left[D_A(\bar z)/r_d\right] }{\left[D_A(\bar z)/r_d\right]_{\rm fid}}$,
where $\bar z=2.34$, the fiducial values $\left[D_H(\bar z)/r_d\right]_{\rm fid}$ and $\left[D_A(\bar z)/r_d\right]_{\rm fid}$ are 8.708 and 11.59 respectively.

\end{itemize}
In the following context, we will use ``$\rm SNLS3$'', ``$\rm Planck$'', ``$\rm BAO$'', ``$\rm HST$'' and ``$\rm SDSS-Ly\alpha$'' to
represent these five datasets.

We combine the above data sets to perform $\chi^2$ analyses.
Since SNLS3, Planck, BAO, HST and SDSS-Ly$\alpha$ are effectively independent measurements,
the total $\chi^2$ function is just the sum of all individual $\chi^2$ functions:
\begin{equation}\label{chitotal}
\chi^2_{\rm total}=\chi^2_{\rm SNLS3}+\chi^2_{\rm Planck}+\chi^2_{\rm BAO}+\chi^2_{\rm HST}+\chi^2_{\rm SDSS-ly\alpha}.
\end{equation}
In our work, for a detailed investigation, we do fittings with two datasets: $\rm Planck+SNLS3+BAO+HST$ and $\rm Planck+SNLS3+BAO+HST+SDSS-Ly\alpha$, respectively.

The $\Lambda$HDE model has two dark energy parameters $d$, and $\Omega_{\Lambda}$.
Including four other cosmological parameters $\Omega_{m}h^2$, $\omega_b$, $\Omega_{k}$ and $h$,
and two nuisance parameters $\alpha$, $\beta$ characterizing the systematic errors of the SNLS3 dataset \citep{SNLS3A},
the full set of free parameters in our analysis is
\begin{equation}\label{Eq:ParSpace}
{\bf P}=\{\Omega_{m}h^2,~\omega_b,~h,~c,~\Omega_{\Lambda},~\Omega_{k},~\alpha,~\beta\}.
\end{equation}
In this work, we numerically solve Eq.(\ref{eq:OH3}) and Eq.(\ref{eq:OH4}) to obtain background evolutions of the  $\Lambda$HDE model.
The values of $\Omega_{r0}$, for simplicity, are
determined from the 7-yr WMAP observations~\cite{WMAP7},
\begin{equation}
\Omega_{r0} = \Omega_{\gamma 0}(1+0.2271 N_{eff}),\ \ \ \Omega_{\gamma 0} = 2.469 \times 10^{-5} h^{-2},\ \ \  N_{eff}=3.046,
\end{equation}
where $\gamma$ represents photons, and $N_{eff}$ is the
effective number of neutrino species.

We modify the public available CosmoMC package~\citep{cosmomc} to explore the parameter space using the Markov Chain Monte Carlo (MCMC) algorithm.
All the parameters listed in Eq. (\ref{Eq:ParSpace}) are fitted simultaneously.

\section{Dynamical Behaviors and the Cosmic Expansion History}

\subsection{Fitting Results}

\begin{table*}
\caption{Fitting results for the $\Lambda$HDE model}

\label{Table1}

\begin{tabular}{cccccc}
\hline\hline &\multicolumn{2}{c}{$\rm Planck+SNLS3+BAO+HST$}&&\multicolumn{2}{c}{$\rm Planck+SNLS3+BAO+HST+SDSS-Ly\alpha$} \\
           \cline{2-3}\cline{5-6}
Parameter  & Best fit & 68.3\% limits & & Best fit & 68.3\% limits \\ \hline
$\Omega_{m}h^2$          & $0.1405$ 
                         & $0.1413^{+0.0025}_{-0.0025}$& 
                         & $0.1375$
                         & $0.1400^{+0.0025}_{-0.0025}$\\ 

$H_{0}$                  & $73.3$
                         & $70.5^{+1.3}_{-1.4}$& 
                         & $71.6$
                         & $70.5^{+1.3}_{-1.4}$\\ 

$d$                      & $0.003$
                         & $0.570^{+0.320}_{-0.180}$&
                         & $0.001$
                         & $0.308^{+0.075}_{-0.308}$\\ 

$\Omega_{\Lambda0}$      & $0.61$
                         & $0.07^{+0.61}_{-0.14}$&
                         & $0.63$
                         & $0.42^{+0.25}_{-0.10}$\\ 

$\Omega_{k0}$            & $-0.0012$
                         & $0.0037^{+0.0037}_{-0.0059}$& 
                         & $-0.0063$
                         & $-0.0013^{+0.0028}_{-0.0032}$\\ 
\hline
$\chi^2_{min}$           ~~~~&   $426.27$  &&
                         ~~&   $431.79$\\ 
\hline
\end{tabular}
\end{table*}

\begin{figure}
\centering{
\includegraphics[height=7.0cm]{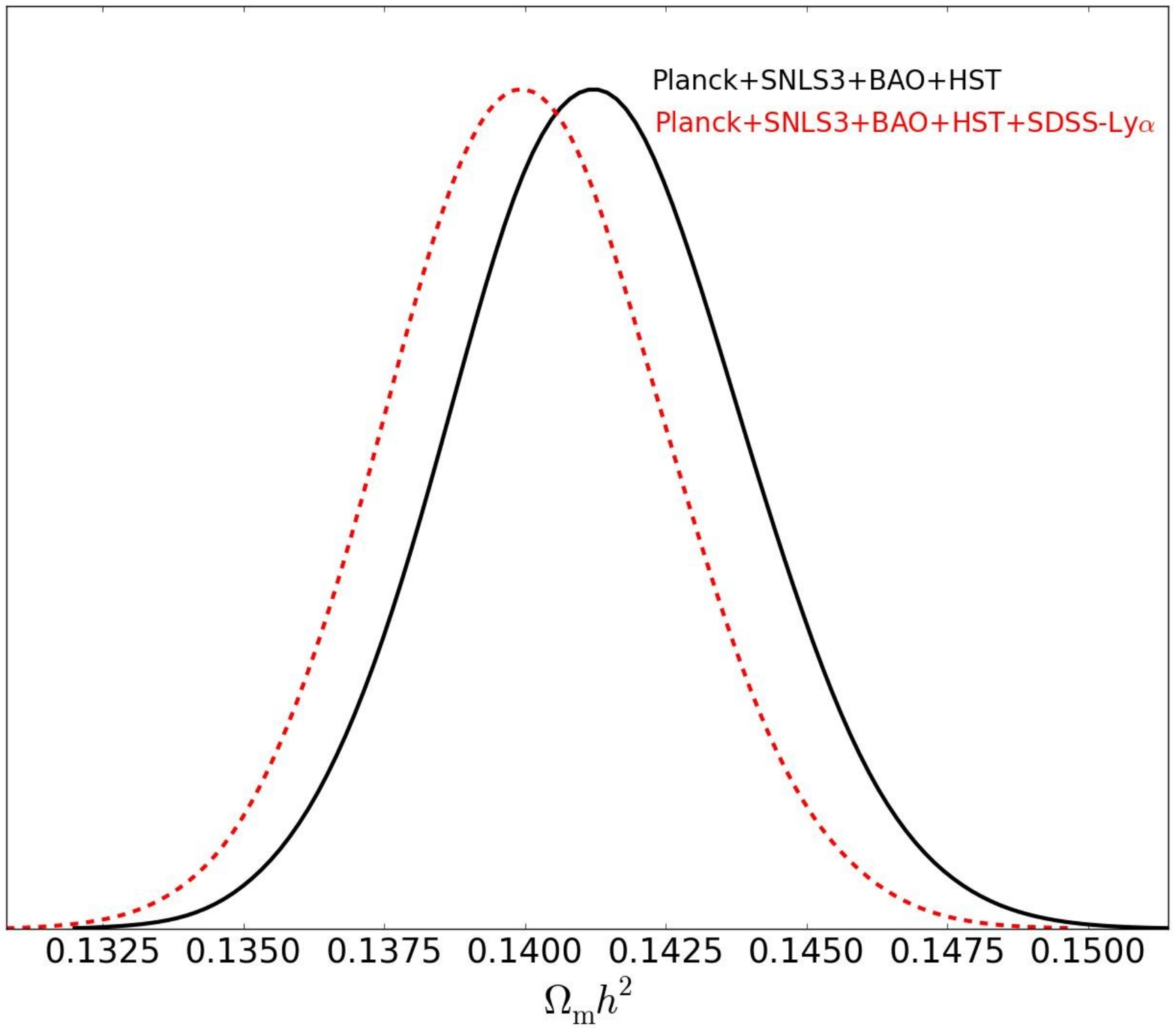}
\includegraphics[height=7.0cm]{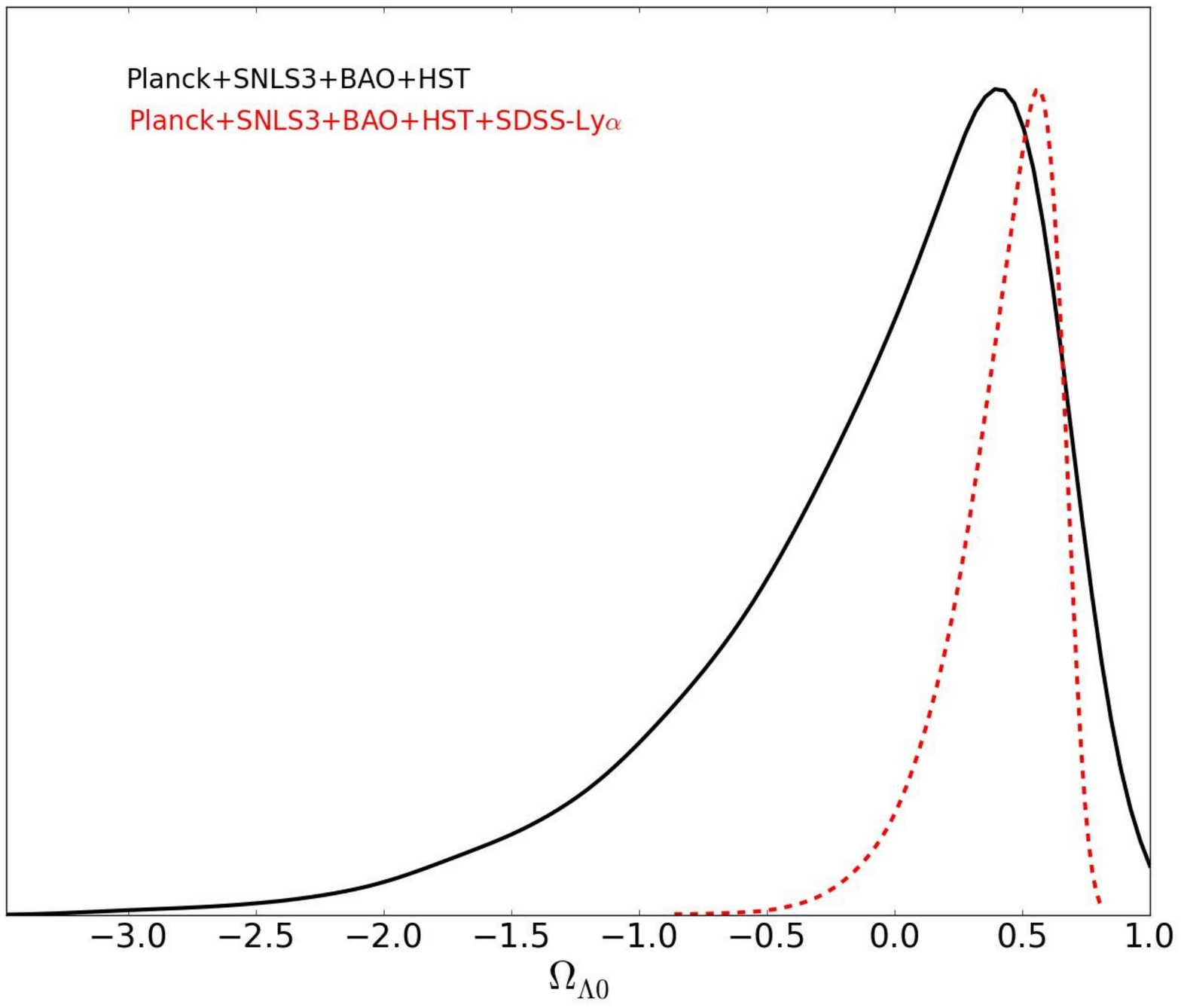}
\includegraphics[height=7.0cm]{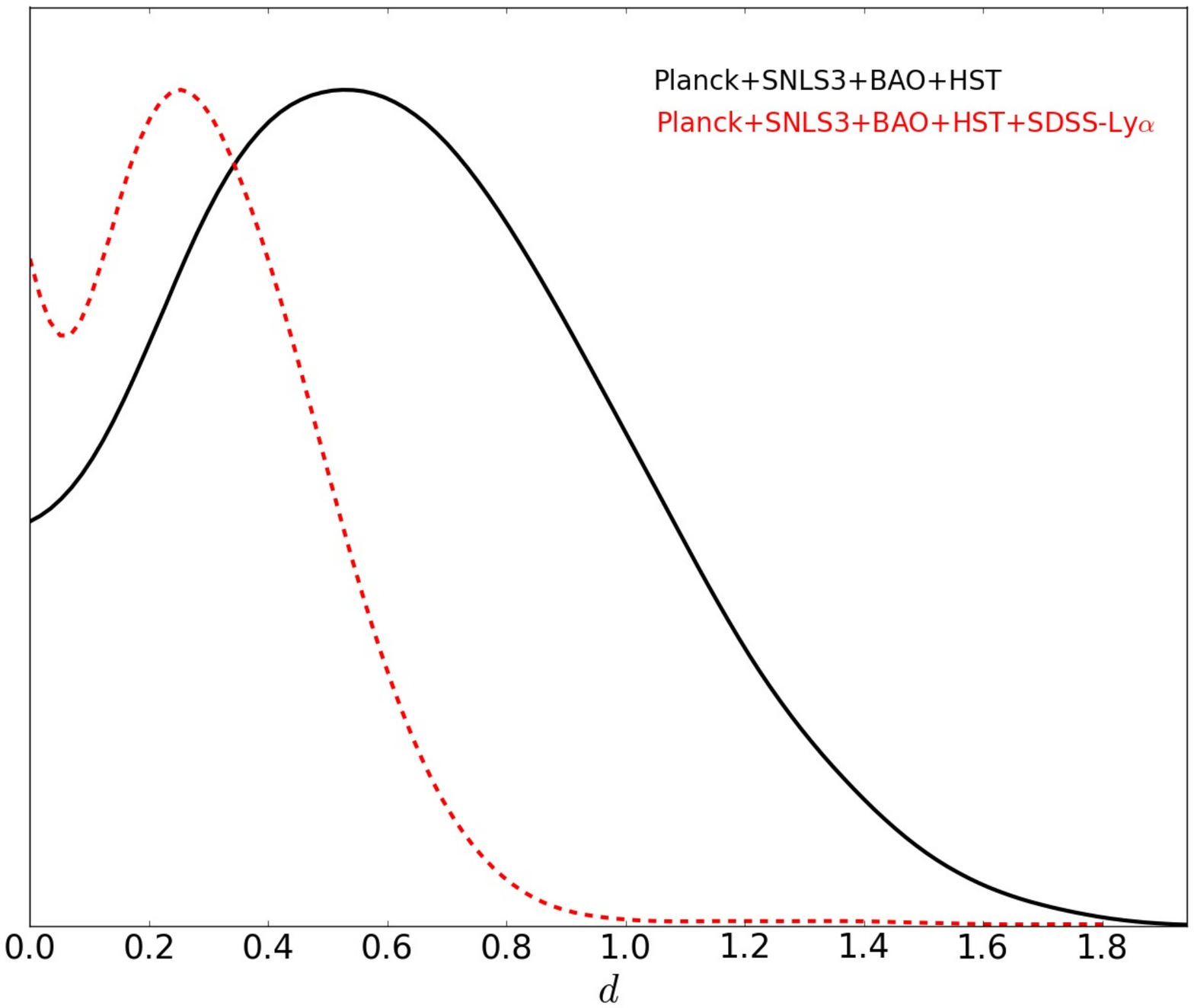}
}
\caption{\label{Fig:Like}
Marginalized likelihood distribution of $\Omega_mh^2$, $\Omega_{\Lambda0}$, $d$ constrained by $\rm Planck+SNLS3+BAO+HST$ (solid lines) and $\rm Planck+SNLS3+BAO+HST+SDSS-Ly\alpha$ (dashed lines) datasets.
}
\end{figure}

\begin{figure}
\centering{
\includegraphics[height=4.0cm]{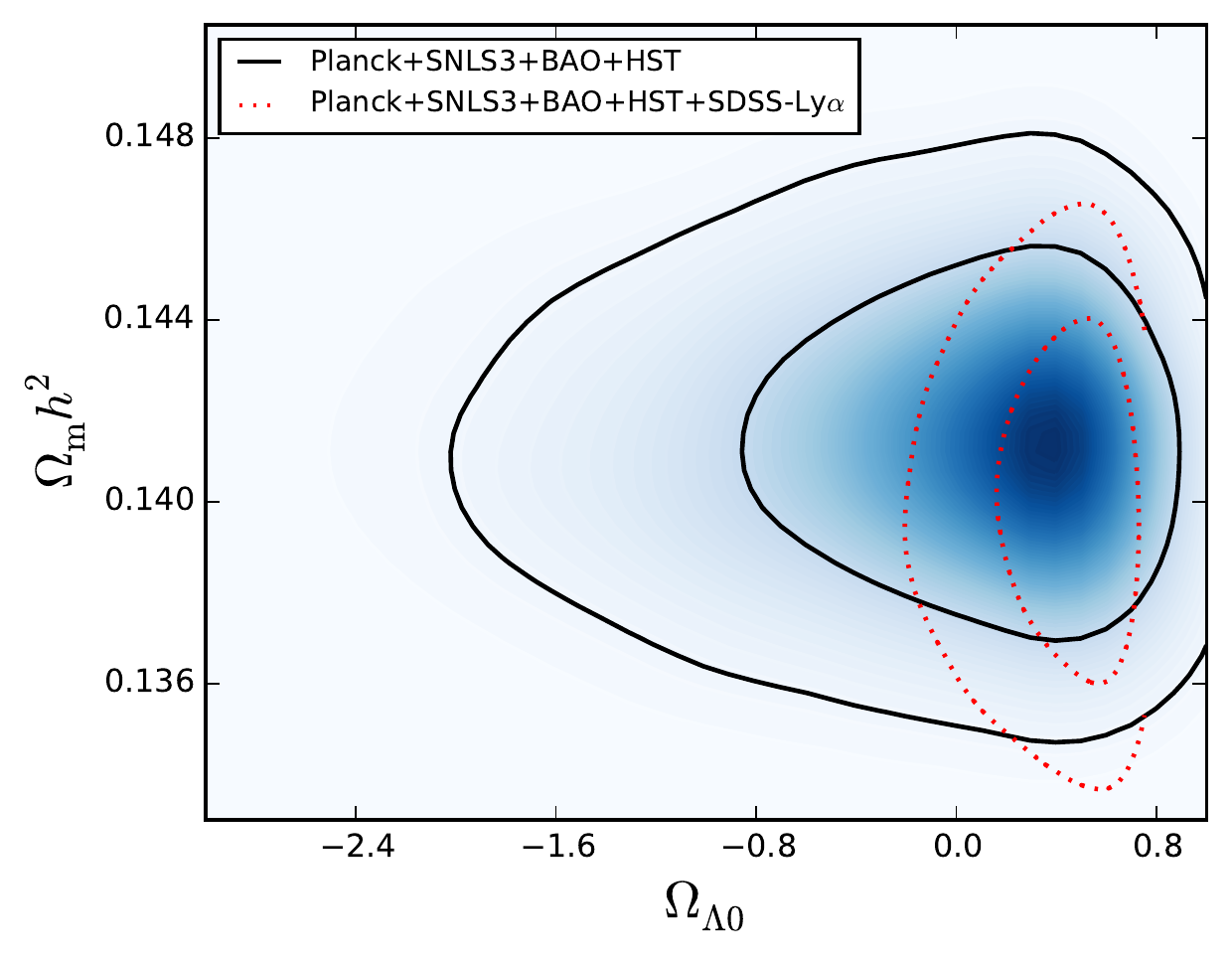}
\includegraphics[height=4.0cm]{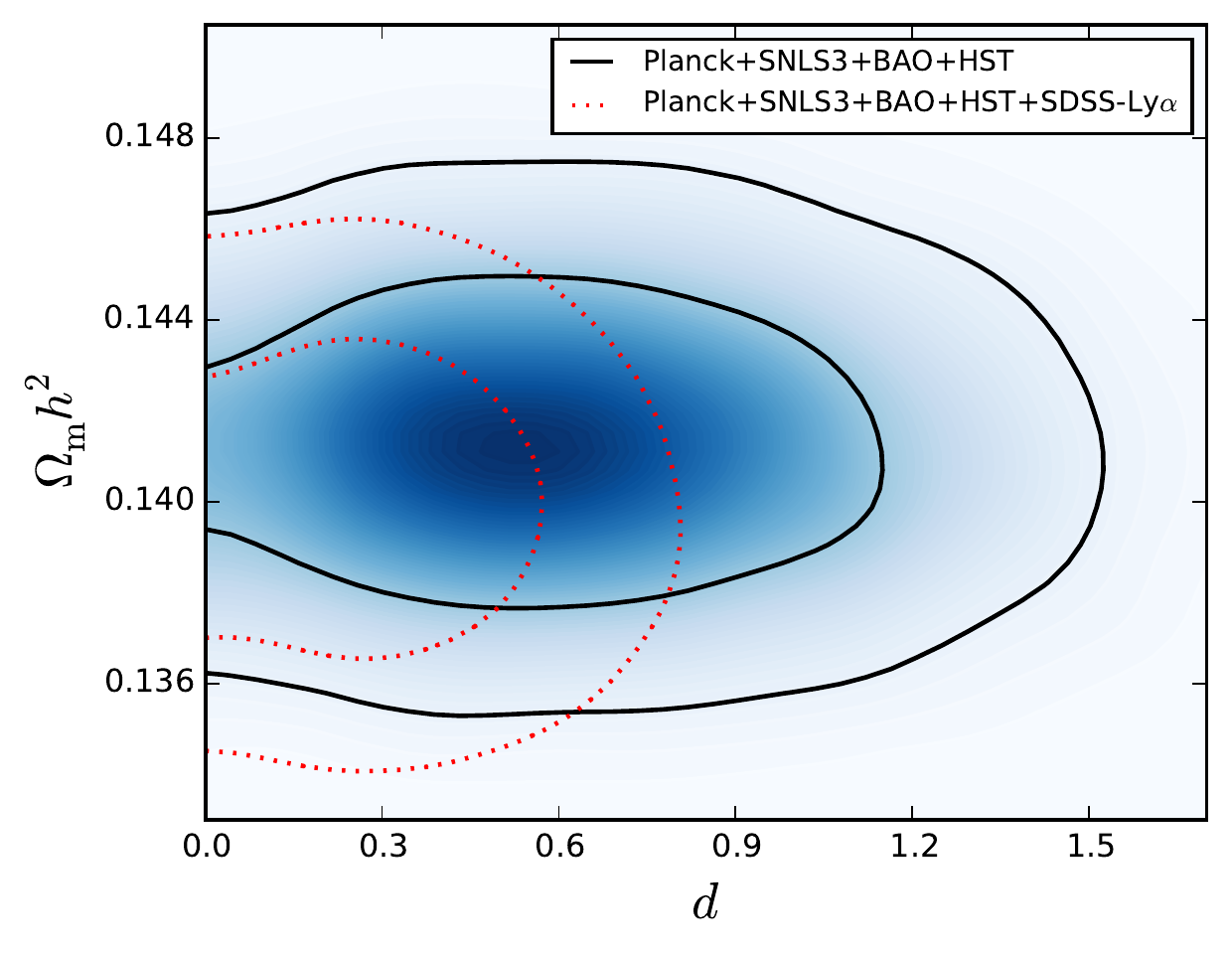}
\includegraphics[height=4.0cm]{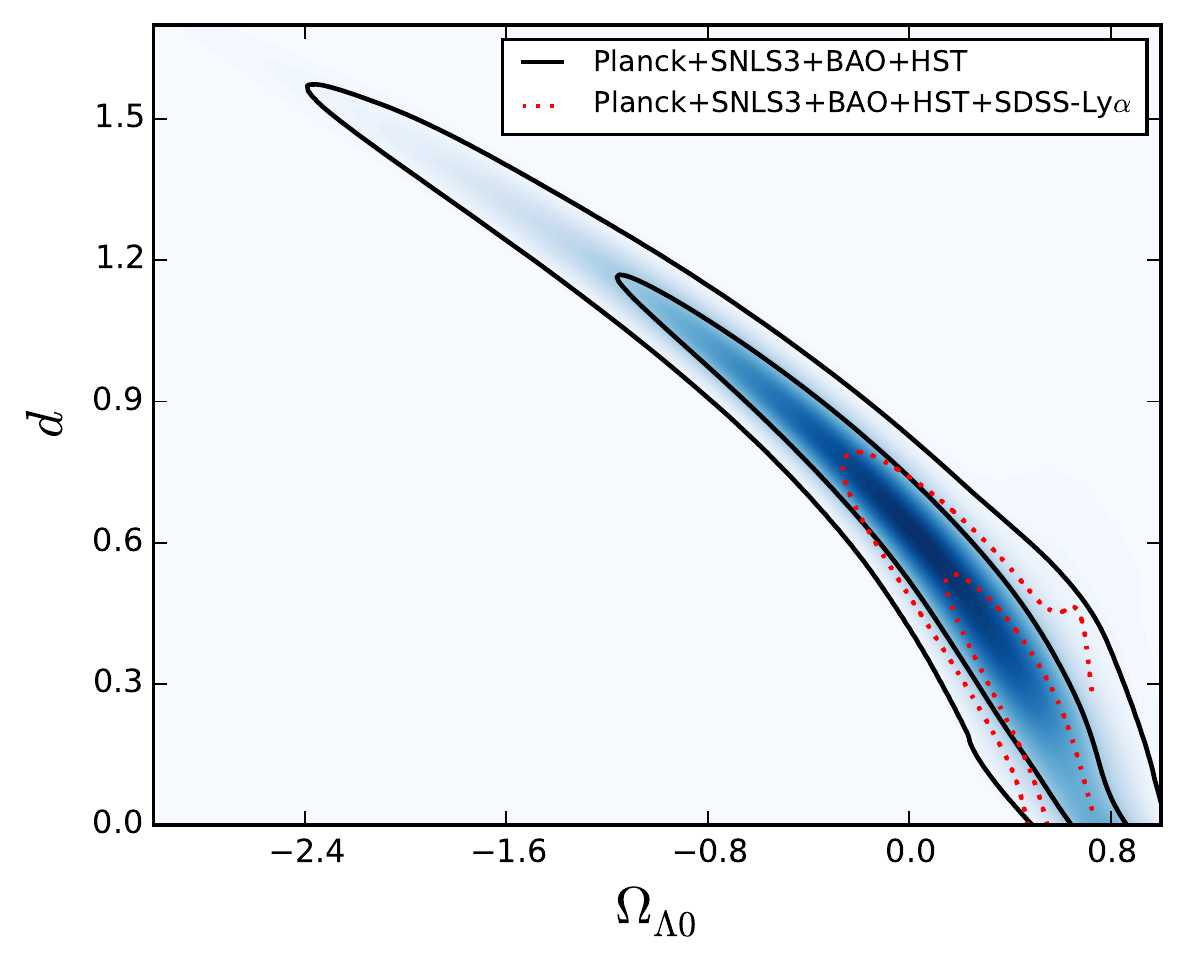}
}
\caption{\label{Fig:Contour}
 Marginalized  68.3\% and 95.4\% CL contours of $\Omega_{\Lambda0}$-$\Omega_mh^2$ , $d$-$\Omega_mh^2$ and $\Omega_{\Lambda0}$-$d$ planes constrained by $\rm Planck+SNLS3+BAO+HST$ (solid lines) and $\rm Planck+SNLS3+BAO+HST+SDSS-Ly\alpha$ (dotted lines) datasets.}
\end{figure}

In Table~\ref{Table1}, we give best-fit parameters as well as 68.3\% confidence limits for constrained parameters.
The results show that the spatial curvatures are close to zero in both cases (the 68.3\% confidence limits are $\left|\Omega_{k0}\right|<0.007$ and $\left|\Omega_{k0}\right|<0.004$, respectively). Thus the results are impressively consistent with a spatially flat universe.
Table \ref{Table1} also gives the constraint on the parameter $\Omega_{\Lambda0}$:

$\rm Planck+SNLS3+BAO+HST$: ~$-0.07<\Omega_{\Lambda0}<0.68$ (68.3\% {\rm CL});

$\rm Planck+SNLS3+BAO+HST+SDSS-Ly\alpha$: ~ $0.32<\Omega_{\Lambda0}<0.67$ (68.3\% {\rm CL}).

Combine with the constraint results of $\Omega_mh^2$ and $H_0$ listed in Table ~\ref{Table1} together, we get the corresponding constraint on $\Omega_{hde0}$:

$\rm Planck+SNLS3+BAO+HST$: ~$0.04<\Omega_{hde0}<0.79$ (68.3\% {\rm CL});

$\rm Planck+SNLS3+BAO+HST+SDSS-Ly\alpha$: ~$0.06<\Omega_{hde0}<0.41$ (68.3\% {\rm CL}).

Fig.~\ref{Fig:Like} shows the marginalized likelihood distributions of parameter  $d$, $\Omega_mh^2$ and $\Omega_{\Lambda0}$ constrained by $\rm Planck+SNLS3+BAO+HST$ and $\rm Planck+SNLS3+BAO+HST+SDSS-Ly\alpha$ datasets, respectively.
The  68.3\% and 95.4\% contours  of $\Omega_{\Lambda0}$-$\Omega_mh^2$ , $d$-$\Omega_mh^2$ and $\Omega_{\Lambda0}$-$d$ planes are plotted in Fig. \ref{Fig:Contour}.

The results show that, compared with the $\rm Planck+SNLS3+BAO+HST$ dataset, the $\rm Planck+SNLS3+BAO+HST+SDSS-Ly\alpha$ dataset makes a more tighter constraints on $d$ and $\Omega_{\Lambda0}$ parameters.
We also find that a smaller value of $\Omega_mh^2$ and a bigger value of $\Omega_{\Lambda0}$ is favored  by the $\rm Planck+SNLS3+BAO+HST+SDSS-Ly\alpha$ dataset.

There is a degeneracy between $d$ and $\Omega_{\Lambda0}$, which can be seen in Fig. \ref{Fig:Contour}.
The reason is that both the cosmological constant and HDE are good candidate for explaining the feature of cosmic acceleration revealed by current observational data.
Therefore, when we combine the HDE and cosmological constant components together (thus $\Lambda$HDE model), we may probably get the degeneracy.

\begin{table*}
\caption{The $\rm \chi^2_{min}$s, $\rm \Delta AIC$s and $\rm \Delta BIC$s of the $\rm \Lambda CDM$, $\rm HDE$ and $\rm \Lambda$HDE models, obtained by using the $\rm Planck+SNLS3+BAO+HST$ and $\rm Planck+SNLS3+BAO+HST+SDSS-Ly\alpha$ datasets, respectively.}

\label{Table2}

\begin{tabular}{cccccccc}
\hline\hline &\multicolumn{3}{c}{$\rm Planck+SNLS3+BAO+HST$}&&\multicolumn{3}{c}{$\rm Planck+SNLS3+BAO+HST+SDSS-Ly\alpha$} \\
           \cline{2-4}\cline{6-8}
Model  & $\rm \chi^2_{min}$ & $\rm \Delta AIC$ & $\rm \Delta BIC$ & & $\rm \chi^2_{min}$ & $\rm \Delta AIC$ & $\rm \Delta BIC$ \\ \hline
$\rm \Lambda CDM$        & $431.35$ 
                         & $0$
                         & $0$& 
                         & $438.22$
                         & $0$
                         & $0$\\ 

$\rm HDE$                & $428.20$
                         & $-1.15$
                         & $3.03$& 
                         & $438.19$
                         & $1.97$
                         & $6.15$\\ 

$\rm \Lambda HDE$        & $426.27$
                         & $-1.06$
                         & $7.27$&
                         & $431.79$
                         & $-2.43$
                         & $5.92$\\ 
\hline
\end{tabular}
\end{table*}

In addition to the cosmological consequence of the $\Lambda$HDE model, we are also interested in its comparison with the $\Lambda$CDM and the original HDE models.
Therefore we also perform the $\chi^2$ analysis of the $\Lambda$CDM and the HDE models by using the same datasets.
To assess different models, here we adopt the Akaike information criteria (AIC) \cite{AIC} and Bayesian information criteria (BIC) \cite{BIC},
defined as
\begin{equation}
  {\rm AIC} = \chi^2_{min}+2k, \ \ {\rm BIC} = \chi^2_{min}+k \ln N,
\end{equation}
where $k$ is the number of free parameters, and $N$ is the number of data points used in the fits.
A model with smaller AIC (BIC) is more favored.

Table~\ref{Table2} shows the $\chi^2_{min}$s, AICs and BICs of the $\Lambda$CDM, HDE and $\Lambda$HDE models.
Notice that the  values of the AIC and BIC themselves are not interesting, thus we only list the difference between the $\Lambda$HDE (HDE) and $\Lambda$CDM models, i.e.,
\begin{equation}
\Delta {\rm AIC}_{\rm model}\equiv {\rm AIC}_{\rm model}-{\rm AIC}_{\rm \Lambda CDM}, \ \ \Delta {\rm BIC}_{\rm model}\equiv {\rm BIC}_{\rm model}-{\rm BIC}_{\rm \Lambda CDM}
\end{equation}
Compared with the $\Lambda$CDM and HDE models, the $\Lambda$HDE model provides a better fit to the data.
For $\rm Planck+SNLS3+BAO+HST$ dataset, the $\Lambda$HDE model reduces the $\chi^2_{min}$s by amount of  5.08 (1.93) compared with  the $\Lambda$CDM (HDE) model.
While for $\rm Planck+SNLS3+BAO+HST+SDSS-Ly\alpha$ dataset, the $\Lambda$HDE model reduces the $\chi^2_{min}$s by amount of about 6.4 compared with the other two models.
By adopting the AIC and BIC, Table~\ref{Table2} also shows that, compared with $\Lambda$CDM model, the $\Lambda$HDE model is slightly favored by AIC.
But both the $\Lambda$HDE and HDE models are not favored by BIC, though these models have slightly smaller $\chi^2_{min}$s than the $\Lambda$CDM model.

\subsection{The Expansion History}

 It is worth investigating the cosmic expansion history of the $\Lambda$HDE model by the fitting results.

\begin{figure}[H]
\centering
\includegraphics[height=7cm]{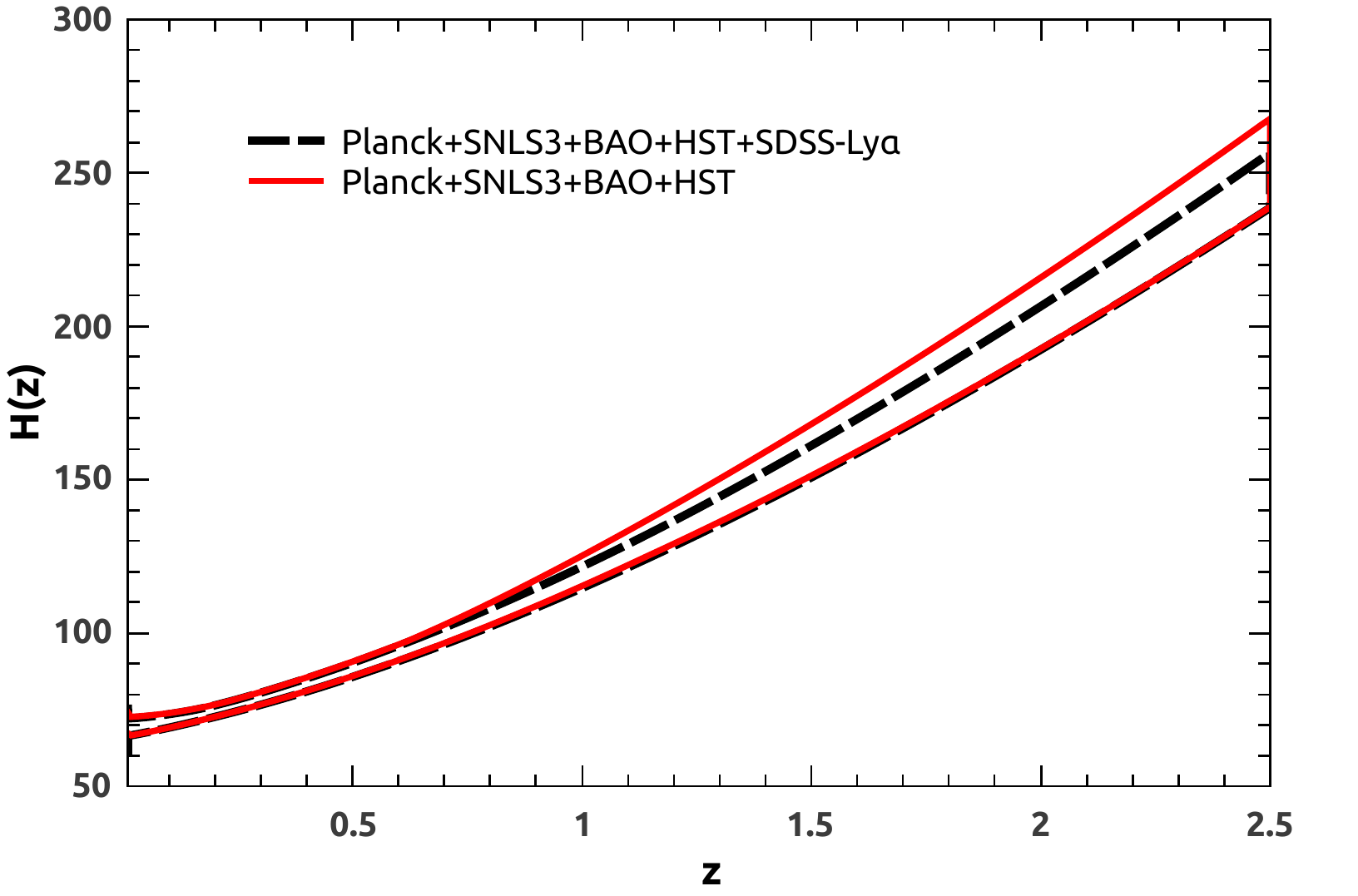}
\caption{\label{Fig:expansionhistory2}
Reconstructed evolution  history of $H(z)$ (95.4\% CL) in $\Lambda$HDE model, constrained by the $\rm Planck+SNLS3+BAO+HST$ (red solid lines) and $\rm Planck+SNLS3+BAO+HST+SDSS-Ly\alpha$ (black dashed lines) datasets respectively.
}
\end{figure}

In Fig. \ref{Fig:expansionhistory2},
we plot the reconstructed evolution  history of $H(z)$ (95.4\% CL) in $\Lambda$HDE model, constrained by the $\rm Planck+SNLS3+BAO+HST$  and $\rm Planck+SNLS3+BAO+HST+SDSS-Ly\alpha$ datasets,  respectively.
We find that, in  low redshift region, the reconstructed evolution history $H(z)$ of  the $\rm Planck+SNLS3+BAO+HST$ and $\rm Planck+SNLS3+BAO+HST+SDSS-Ly\alpha$ datasets are almost the same.
However, in the high redshift region, the constraint of the $\rm Planck+SNLS3+BAO+HST+SDSS-Ly\alpha$ dataset is much more  tighter than that of the $\rm Planck+SNLS3+BAO+HST$ dataset. It is clear that this feature is due mainly to the SDSS-Ly$\alpha$ data at redshift $z=2.34$. As revealed by \cite{sahni2014,Yazhou2014}, the high redshift datasets play a big part in constraining the cosmic expansion history.

\begin{figure}[H]
\centering
\includegraphics[height=7cm]{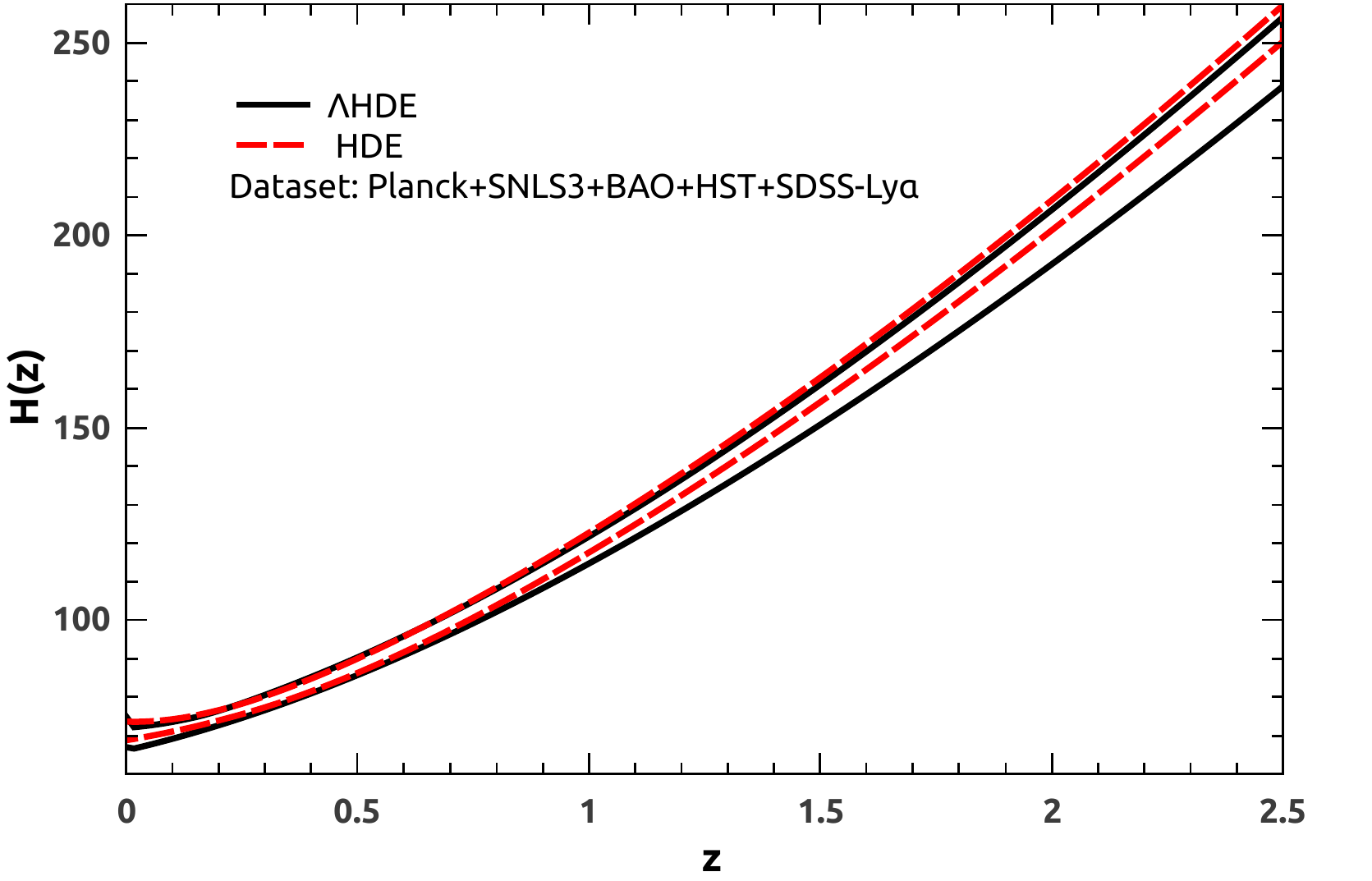}
\caption{\label{Fig:expansionhistory1}
Reconstructed evolution history of $H(z)$ (95.4\% CL) constrained by $\rm Planck+SNLS3+BAO+HST+SDSS-Ly\alpha$ dataset in $\Lambda$HDE model(black solid lines) .
The corresponding results for HDE model is also plotted (red dashed lines) for comparison.}
\end{figure}

For a comparison, it is also of value to compare the $\Lambda$HDE model with the original HDE  model.  Fig. \ref{Fig:expansionhistory1} shows the reconstructed evolution  history of $H(z)$ (95.4\% CL) in the $\Lambda$HDE and the original HDE model constrained by $\rm Planck+SNLS3+BAO+HST+SDSS-Ly\alpha$ dataset.
We find that the reconstructed $H(z)$ of $\Lambda$HDE and HDE models have negligible difference in the low redshift region.
However, in the high redshift region, the $H(z)$ in $\Lambda$HDE model has slightly lower value,
which should be due mainly to the existence of a cosmological constant component in the model.

\subsection{Equation of State}

In this subsection we discuss the EoS $w$,
which is believed to be the most important marker of the properties of dark energy.

\begin{figure}[H]
\centering{
\includegraphics[height=7cm]{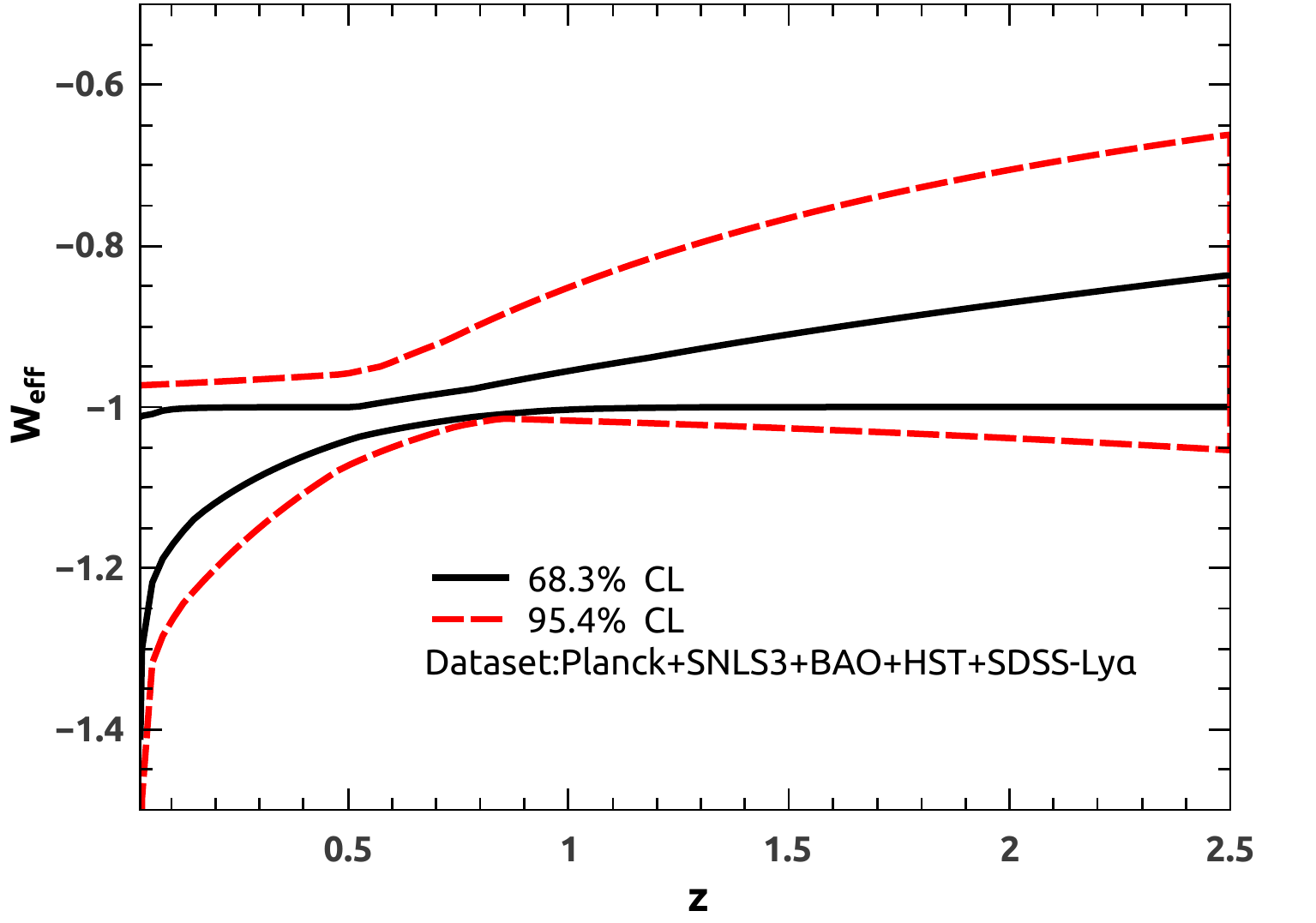}}
\caption{\label{Fig:wz}
Reconstructed evolution history of $w(z)$  at 68.3\% (black solid lines) and 95.4\% (red dashed lines) CL in $\Lambda$HDE model, constrained by the $\rm Planck+SNLS3+BAO+HST+SDSS-Ly\alpha$  dataset.}
\end{figure}

Fig. \ref{Fig:wz} shows the reconstructed evolution  history of $w(z)$ at $0\leq z \leq 2.5$(68.3\% and 95.4\% CL) constrained by $\rm Planck+SNLS3+BAO+HST+SDSS-Ly\alpha$ dataset.
It shows that $w$ slightly cross -1 from above roughly at the current epoch.
However, in the past we have $w$ slightly bigger than -1,
which can be viewed as a feature of diluted holographic dark energy.
This behavior is consistent with the results shown by Fig. \ref{Fig:expansionhistory1}.

As mentioned above, the dynamical evolution of dark energy have not be confirmed
by the current observational data. Our results is consistent with this statement.

From Fig. \ref{Fig:expansionhistory2}, we can conclude that, if we want to break the degeneracy between the HDE and cosmological constant components, one way is to get more observational data at high redshifts.

\section{Conclusion}

In this work,  we study  the $\Lambda$HDE model in which there are two parts, the first being the cosmological constant and the second being the holographic dark energy.
This model has similar dynamical equations with the original HDE model,
except a cosmological term. By studying the $\Lambda$HDE model theoretically, we find that the parameters $d$ and $\Omega_{hde}$
are divided into a few domains in which the fate of the universe is quite different.

Using the $\rm Planck+BAO+SNLS3+HST$ and $\rm Planck+BAO+SNLS3+SDSS-Ly\alpha$ datasets,
we investigate the dynamical properties and cosmic expansion
history of the $\Lambda$HDE model.
The results shows that the goodness-of-fit of the $\Lambda$HDE model
are $\chi^2_{\rm min}$=426.27 ($\rm Planck+SNLS3+BAO+HST$)
and $\chi^2_{\rm min}$=431.79 ($\rm Planck+SNLS3+BAO+HST+SDSS-Ly\alpha$)
which is smaller then the results of the original HDE model
($\rm Planck+SNLS3+BAO+HST$:428.20; $\rm Planck+SNLS3+BAO+HST+SDSS-Ly\alpha$:438.19)
and the concordant $\Lambda$CDM model ($\rm Planck+BAO+SNLS3+HST$:431.35;
$\rm Planck+SNLS3+BAO+HST+SDSS-Ly\alpha$:438.22) obtained using the same datasets. Especially when constrained by  the $\rm Planck+SNLS3+BAO+HST+SDSS-Ly\alpha$ dataset, The $\chi^2_{\rm min}$ of $\Lambda$HDE model shrinks more than 6,  compared with both the HDE and $\Lambda$HDE model.
Thus, the $\Lambda$HDE model provides a nice fit to the cosmological data.

For parameter $\Omega_{\Lambda0}$, the 68.3\% confidence level constrained by the $\rm Planck+SNLS3+BAO+HST$
and the $\rm Planck+SNLS3+BAO+HST+SDSS-Ly\alpha$ dataset is
$-0.07<\Omega_{\Lambda0}<0.68$ and $0.32<\Omega_{\Lambda0}<0.67$, respectively.
This gives the corresponding components of the holographical dark energy, namely

$\rm Planck+BAO+SNLS3+HST$: ~$0.04<\Omega_{hde0}<0.79$;

$\rm Planck+SNLS3+BAO+HST+SDSS-Ly\alpha$: ~$0.06<\Omega_{hde0}<0.41$.

We also find that there is degeneracy between the cosmological constant and the holographic dark energy component when constrained by current cosmological observations.
By reconstructing the evolution of the EoS of dark energy,
we find the $\Lambda$HDE mainly differs from the original HDE model at high redshift (as shown in Fig.\ref{Fig:expansionhistory1}).

From the constraint results by $\rm Planck+SNLS3+BAO+HST+SDSS-Ly\alpha$ dataset, it shows that if we want to break the degeneracy between the HDE and cosmological constant components, one way is to get more observational data at high redshifts.

\acknowledgments

We are grateful to the Referee for the valuable suggestions.
We also thank Xiao-Dong Li and Shuang Wang for their helps on fitting issues.
ML is supported by the National Natural Science Foundation of China (Grant No. 11275247, and Grant No. 11335012) and 985 grant at Sun Yat-Sen University.
\

\

\

\

\

\
\section{Appendix: Proof for $\log a$ not having a maximum in $\Lambda$HDE model}\label{Data}
In this appendix, we prove that $x$ does not have a maximum as a function of time.
Since $x$ does not have a maximum in either $\Lambda$CDM or HDE model, here we only consider $\Lambda$HDE model, which includes the coexistence case of  HDE and cosmological constant.

In a flat universe, the Friedmann equation is
\begin{equation}
\label{eq:FE2} 3M_{pl}^2 H^2=\rho_{dm}+\rho_b+\rho_r+\rho_{\Lambda}+\rho_{hde},
\end{equation}
Let us note that
\begin{eqnarray}
  g(x)&=&\frac{d}{dx}\ln{\left|f(a)\right|},
  \\
{\rm then} \quad g(x)&=&\frac{2\Omega_{\Lambda0}e^{2x}-\Omega_{m0}e^{-x}-2\Omega_{r0}e^{-2x}}
  {\Omega_{\Lambda0}e^{2x}+\Omega_{m0}e^{-x}+\Omega_{r0}e^{-2x}}.
\end{eqnarray}
According to Eq.(\ref{main}), we get
\begin{equation}~\label{omega_hde}
\frac{d\Omega_{hde}}{dx}=[\frac{2}{d}\sqrt{\Omega_{hde}}-g(x)]\Omega_{hde}(1-\Omega_{hde}).
\end{equation}

We first consider the cases that $\Omega_{\Lambda0}>0$ or $\Omega_{\Lambda0}+\Omega_{m0}+\Omega_{r0}<0$. In these cases, $g(x)$ is a bounded function in the region ($x_0, +\infty$).
If $x$ has a maximum $x_m$, $H$ would approach zero when $x$ goes to $x_m$.
According to Eq.(\ref{eq:FE2}), $\Omega_{hde}$ would approach infinity.
However, Eq.(\ref{omega_hde}) shows that, once $\sqrt{\Omega_{hde}}>\frac{d}{2}g(x)$ and $\Omega_{hde}>1$, $\frac{d}{dx}\sqrt{\Omega_{hde}}<0$. Thus $\Omega_{hde}$ would never approach infinity.
So $x$ has no maximum in these cases.

Another case is $\Omega_{\Lambda0}<0$ and $\Omega_{\Lambda0}+\Omega_{m0}+\Omega_{r0}>0$.
In this case
\begin{equation}
\exists x_{c}>0 :\quad
\Omega_{\Lambda0}e^{2x_{c}}+\Omega_{m0}e^{-x_{c}}+\Omega_{r0}e^{-2x_{c}}=0.
\end{equation}
So Eq.(\ref{omega_hde}) has a singularity at $x=x_c$.
Similar to the previous analysis, $x$ does not have a maximum in the region ($x_0$, $x_c$).
Now we prove that $x_c$ is not a maximum of $x$.
If it is, $g(x)$ would approach negative infinity and $\Omega_{hde}$ would approach infinity when $x$ goes to $x_c$.
However, once $\Omega_{hde}>1$, $\frac{d}{dx}\sqrt{\Omega_{hde}}<0$.
Thus $\Omega_{hde}$ would never approach infinity.
So $x_c$ is not a maximum of $x$.
When $x>x_c$, we can see that $\Omega_{\Lambda}+\Omega_{m}+\Omega_{r}<0$.
This is just the case we have discussed.
So $x$ also has not a maximum when $x>x_c$.

In summary, we have proved that $x$ does not have a maximum.
So we can use a constant $k_r$ to approximate $\frac{d}{dx}\ln{\left|f(a)\right|}$ when we study the fate of the universe.



\end{document}